\documentclass[aps,showpacs,preprintnumbers,amsmath,amssymb,nofootinbib]{revtex4}

\usepackage{graphicx}
\usepackage{color}

\def \be  {\begin{equation}}
\def \ee  {\end{equation}}
\def \ee  {\end{equation}}
\def \bea {\begin{eqnarray}}
\def \eea {\end{eqnarray}}

\def\be {\begin{equation}}
\def\ee {\end{equation}}
\def\bea {\begin{eqnarray}}
\def\eea {\end{eqnarray}}
\def\bc {\begin{center}}
\def\ec {\end{center}}
\def\bfg {\begin{figure}}
\def\efg {\end{figure}}
\def\bi {\begin{itemize}}
\def\ei {\end{itemize}}

\def\beq{\begin{equation}}
\def\eeq{\end{equation}}
\def\br{\begin{eqnarray}}
\def\er{\end{eqnarray}}
\newcommand{\eel}[1] {\label{#1}\end{equation}}
\begin{document}
\def\appendixa{
 \vskip 1cm
 {\bf APPENDIX A:kinetic energy}
 \vskip 1cm
 \par
 \setcounter{equation}{0}
 \def\theequation{A.\arabic{equation}}
 }
 \def\appendixb{
\vskip 1cm
 {\bf APPENDIX B: Wave function with the Clebsch-Gordon coefficient}
 \vskip 1cm
 \par
 \setcounter{equation}{0}
 \def\theequation{B.\arabic{equation}}
 }
%
\preprint{2020-3}
\title{Study of the Ground State Energies of Some Nuclei Using Hybrid Model}

\author{R.Hussien}
\email{rabab.hussien216@gmail.com}
\email{rabab.asar17@fsc.bu.edu.eg}
\affiliation{Mathematics :: Theoretical Physics Department, Atomic Energy Authority, Cairo, Egypt}
\author{Sh.M.Sewailem}
\email{Sh_m_sw@yahoo.com}
\affiliation{Mathematics :: Theoretical Physics Department, Atomic Energy Authority, Cairo, Egypt}

\author{L. I. Abou-Salem}
\email{loutfy.Abousalem@fsc.bu.edu.eg}
\affiliation{Physics Department, Faculty of Science, Benha University, Benha, Egypt}

%
%
%
\begin{abstract}
The quark-quark $QQ$ interaction as a perturbed term to the nucleon-nucleon interaction $NN$ without any coupling between them is studied in a hybrid model. This model is used to calculate the ground-state energies of $^{2}H_{1}$ and $^{4}He_{2}$ nuclei. In a semi-relativistic framework, this model is encouraged for light nuclei and the instanton induced interaction by using the QQ potential and the NN interaction for a small scale around the hadron boundaries.This hybrid model depends on two theories, the one-boson exchange potential $OBEP$ and the Cornell-dressed potential $CDP$ for $QQ$. A small effect of quark-quark interaction is obtained on the values of the ground state energies, around $6.7$ for $^{2}H_{1}$ and $1.2$ for $^{4}He_{2}$ by using the considered hybrid model.

\end{abstract}
\maketitle
\section{Introduction}
One of the fundamental problems of the nuclear structure is the derivation of the ground state energies through different methods. Such as properties related to the constituents of matter, which are represented in the physics of elementary particles with their characteristics and how each particle interacts with others. The interaction between each nucleon with all other nucleons generates an average potential field where each nucleon moves. The rules of the Pauli exclusion principle govern the occupation of orbital quantum states in the shell model and postulate that under the meson exchange between two nucleons, the wave function is the anti-symmetrical product wave function. The calculation of the nuclear mean-field potential with Dirac-Hartree–Fock qualifies the description of nucleon-nucleon interaction to be successful microscopically. The interaction between two nucleons has three regions with three ranges. The first region originated from Pseudo scalar meson, the second region related to the scalar meson, and the third region caused by the exchange of vector meson besides the effects of quantum chromodynamics $(QCD)$. The nucleon-nucleon potential has no definite method to determine it, The Bonn group potential known as one-boson exchange potential is supposed to be the suitable model for this interaction because of the reduction of free parameters and fitting them accurately with the experimental data.

\par
On the other hand, the quark degrees of freedom are under the dynamics of $QCD$. The interaction between quarks has various forms of potentials, these forms have to regard the quark properties (confinement and asymptotic properties). The mechanism of the one-gluon exchange approach is dominant at the short-range with two parts. The linear confinement at a long-distance and a part of the asymptotic property represented in the pairing force acting only on the quark-antiquark states. The constituents of baryons composed of $u, d, s$ quarks can use a semi relativistic potential model that refers to their interaction, including the instanton induced forces. The instanton-induced model is used to describe baryons composed of light quarks that are demanded in the considered baryons. This interaction resembles the tunneling phenomena as it can be affected outside the hadron for a short scale comparing with the confinement scale. In the used model we have two contributions in the potential form, the one-gluon exchange part and the exchange of pseudo particles between quark-antiquark pair. The possibility of proposing a hybrid model with no coupling between quarks inside the baryon and mesons. The possibility of proposing a hybrid model with no coupling between quarks inside the baryon and mesons outside it can be founded based on the variational concept of physics. In the present work, The ground state energies for some light nuclei can be calculated successfully by using the considered hybrid model.

In section II, we introduce the theory of $NN$ interactionthrough the one-boson exchange potential with investigations and motivations of the formula. In section III, we have a brief look at the $QQ$ interaction, and the reason for choosing Cornell dressed potential is mentioned to make the idea of the hybrid model possible. In section IV, where the theoretical analysis for the construction of one-boson exchange potential through the exchange of two, three, and four mesons is clarified. Section V shows the theoretical analysis of $QQ$ interaction and the final form of $CDP$. Finally, in sections VI and VII, the obtained results and conclusion are given.

 \section{Theory of $NN$ through the $OBEP$ }
The start of using the fact that there is no unique potential for determination of the effective $NN$ potential, leads to exist different forms with different methods.
So, this work concerned to show the effect of our potential which is published in previous work \cite{rabab}. Our potential is constructed with the idea of one-boson exchange and also depends on the motion of nucleons in the nucleus. This motion produced a field called nuclear mean field, the interaction between nucleons is controlled with Pauli principle and nuclear shell-model. We considered the spatial exchange between two nucleons so, the non-local field is determined with Hartree-Fock approximation. Since the Fock effect is demonstrated in the non-vanishing spatial components for the vector part of the potential. Our potential implies Dirac-Hartree-Fock method to determine the wave function and energy of a quantum N-body system in a stationary state. We classified our potential as a semi-relativistic model because of neglecting the fourth power of momentum to simplify the formula and it will be included in following work. Our potential is associated with Bonn group to have the meson's function and its parameters. To calculate the ground state energies of Hartree-Fock approximation, we need to minimize the total energy of single particle potential by the Steepest descent methods directly to have the lowest energy. It is demanded a modification for nucleons wave functions and energy.\\
The modification of wave function is demonstrated by Clebsch-Gordon coefficients, and Talmi-Moshinsky harmonic oscillator bracket, affecting on the radial, spin and isotopic wave functions. We use the formalism of second quantization just as a convenient way of handling antisymmetric wave function. This formalism referred as a representation of the occupation number, hence it leads to be represented in the Fock-state basis which can be constructed by filling up each single-particle state with a certain number of identical particles. As a real space basis, we write the antisymmetric wave functions in a Slater-determinant. Second quantization give us the ability to displace the wave function as a Dirac-state and do the same for Slater-determinant. So, we use operators to specify the occupied orbitals and the field operators to define the coordinates for the real space representation. It is noticed that, the atom in a quantum state of energy $E$ depends only on that energy through the Boltzmann factor and not on any other property of the state when we represent this atom in complete thermal equilibrium to determine the ground state energy for the considered nuclei. We use this fact to neglect the tensor force for the Deuteron nucleus and calculate its wave function in S-state only.\\
The fact of being the vector mesons and $QCD$ affected on the nuclear properties at short range, hence the studying of nucleon-nucleon interaction through $OBEP$ should not be enough. The exchange of bosons with $OBE$ potential models comes about more than size of nucleon or equal to the inter-nucleon distances. We have the effect of $QCD$ dynamics at distance less than or almost around the boundaries of hadron and that is necessary for the description of nucleon-nucleon interaction. The quantitative theoretical models can analyze the experimental data based on the degrees of hadrons over the last three decades \cite{Machleidt,tourreil,nagels,erkelenz,holinde,fleischer} and also the quark degrees of freedom in $QCD$ models are a successful models for the description of the nuclear properties \cite{lucha,plante,kang,su}. These models analyze the static properties of baryon successfully. We are concerned to add the quark degrees of freedom as a perturbed term to the meson degrees of freedom and have a hamiltonian equation of two parts as following.
\begin{equation}\label{hamiltonian}
  H=H_{NN}+H_{QQ}
\end{equation}
Where the hamiltonian of the nucleon-nucleon interaction is $H_{NN}$ and the hamiltonian of the quark-quark interaction is $H_{QQ}$. So, we study the $OBEP$ with the exchange of three and four mesons as it represents  the nucleon-nucleon interaction, and the Cornell dressed potential as the quark-quark interaction for constructing more realistic model of the nuclear properties.

 \section{Theory of $QQ$ through the $CDP$}
The simulation of quark-quark interaction phenomenon in a semi-relativistic frame work shows that the long range part of this interaction increasing linearly with the distance and called it confinement, and the short range of interaction is a result of Coulomb-like interaction (one-gluon exchange).
The idea of considering the contribution of the constant potential is dominant than the other contributions to the quark-quark potential worth good thinking of it as in \cite{brausemay,cfb}, and a good spectra of mesons and baryons are obtained. At first, very good results for the charmonium spectrum obtained from a simple form of a potential called Funnel potential or Cornell potential \cite{blask,resag}.
The Hamiltonian of hadrons containing light quark should simultaneously define a number of relativistic corrections. The momentum-dependent corrections as well as a non-local kinetic energy (there is no commutation with faster than light and it is compatible with special relativity) are required to be included in the effects of relativistic kinematics of the potential of the potential energy operator \cite{fulcher}. These relativistic kinematics are included in the Bethe-Salpter equation, neglecting the spin effect which introduce non-local modifications of the relative coordinate. The spinless Salpter equation has the form,
\begin{equation}\label{hamiltonian quark}
  H=\sum^{3}_{i=1}\sqrt{\vec{p_{i}}^{2}+m_{i}^{2}}+\sum^{3}_{i<j=1}V_{ij}
\end{equation}
This is suitable for baryon where $H$ is the total energy of the system, $V$ is the central potential between two particles $(i,j)$ and $\vec{p}$ is their relative momentum. In case of baryon $m_{i}$ is the constituent masses of quarks with the same mass of $u$ and $d$ quarks (isospin symmetry is maintained). The central potential is the Cornell-potential,
\begin{equation}\label{cornell}
  V_{C}(r)=\frac{1}{2}[\frac{-k}{r}+ar+C]
\end{equation}
The factor half is related to the half rule, $k$ is the Coulomb parameter, $a$ is the string constant and $C$ is additive constant equals zero in the heavy quark sector. To solve the spinless-Salpter equation, \cite{fulcher} expanded the wave function in terms of a complete set of basis functions according to Rayleigh, Ritz and Galerkin method as the previous radial wave function in $OBEP$.
Many symmetries are slightly broken in nature as it can give rise the classical solutions to a particular symmetry-breaking amplitude. This amplitude is similar to the tunneling effect, indeed the classical solutions of the equation of motion can sometimes describe the tunneling through a barrier. The classical solution of equations of motion was introduced in Yang-Mills theory, known as 'instanton' term. The computation of the quantum effects of instantons was introduced by 't Hooft \cite{hooft} firstly. In \cite{brausemay,cfb}, the authors went with instanton induced interaction (it is a solution to the equations of motion of the classical field theory, it is supposed to be critical points of the action for such quantum theories), In non-relativistic quark model, it is assumed that this model is based on the confinement potential and a residual interaction. The residual interaction is related to the reduction of the one-gluon exchange $OGE$. One able to compute the residual interaction by 't Hooft force from instanton effects \cite{hooft,shifman}. Ref \cite{blask} proposed a model of quarks interaction with the replacement of the traditional $OGE$ potential by a non-relativistic limit of 't Hooft's interaction. The residual interaction is observed by 't Hooft as an expansion of the Euclidean action around the single instanton solutions under the assumption of zero mode in the fermion sector. This interaction has an effective lagrangian with effective potential between two quarks. The instanton calculus can be summarized by four steps \cite{grafke}.
\begin{itemize}
  \item The gluon fields can not deform the instantons into classical solutions continuously.
  \item The perturbative gluon diagrams can not cover the effective interaction between quarks which caused by instantons.
  \item The instanton calculus denotes as a non-perturbative method for the calculation of path integrals, which are represented in the fluctuations around the instanton and change the action. All of this is normally done in the Gaussian approximation.
  \item The instanton effects in $QCD$ realized that instanton is similar to be described as $4$-dimensional gas of pseudo particles, then use the summation over the instanton gas.
\end{itemize}
In the first analogy of super conductivity with the Bardee-Cooper-Schrieffer theory \cite{cooper}. When the interaction between fermions(nucleons) and light quarks are attracted strongly at the short range, this interaction can rearrange the vacuum and the ground state affected by it which resembles the effect of super conductivity. Then, the short range interaction can bind these constituent light quarks into hadrons without confinement in order to make quantitative predictions for hadronic observable. It is clarified that instanton is represented as a tunneling event between vacua \cite{schafer}.\\
 \section{Theoretical analysis of $OBEP$}
The general form which describe the ground state energy of the considered system is the following.
\begin{equation}\label{general}
 H|\Psi\rangle=E|\Psi\rangle
 \end{equation}
Where $H$ is the Hamiltonian and $E$ is the total energy of the system.
 \begin{equation*}
 E = T+V
 \end{equation*}
Defining $T$ is kinetic energy and $V$ is the potential energy. Hence the Hamiltonian of fermions interacting via the potential $V_{ij}$. Thus the accurate Hamiltonian interaction of the nuclear system can be described by Dirac to represent the number of fermions s' interaction where this Hamiltonian is \cite{Miller, Jaminan80, Arias,Meibner},
 \begin{equation}\label{1}
H=\sum_{i}^{A}c\vec{\alpha_{i}}.\vec{p_{i}}+(\beta_{i}-I)m_{i}c^2+T_{ij}+\frac{1}{2}\sum_{i\neq j}^{A}V_{ij}
\end{equation}
Since $\vec{\alpha}$ and $\beta$ are $4\times4$ Dirac matrices, $c$ is the speed of light, $m_{i}$ is the nucleon mass, $T_{ij}$ is the relative kinetic energy and $\vec{p}$ is the momentum operator.\\
See \textbf{appendix $A$} for the details of relative kinetic energy calculations to have the following equation \cite{Neff,Myo,Zhang,Gartenhaus}.
 \begin{equation}\label{4}
T_{ij}=\frac{2}{m A}\sum_{i<j}p_{ij}^{2}
 \end{equation}
Substituting the last equation in Eq (\ref{1}), thus the relativistic Hamiltonian operator for bound nucleons which interact strongly through the potential can be expressed as following.
\begin{equation}\label{r1}
H=\sum_{i}^{A}C\vec{\alpha_{i}}.\vec{p_{i}}+(\beta_{i}-I)m_{i}c^2+\frac{2}{m A}\sum_{i<j}^{A}p_{ij}^{2}+\sum_{i<j}^{A}V_{ij}
\end{equation}
In Hartree-Fock theory, we seek for the best state giving the lowest energy expectation value of this hamiltonian to determine the ground state energy of the considered nuclei.
One able to ensure the antisymmetry of the fermions' wave functions with the aid of Slater Determinant introduced in(1929) and Hartree product to have the convenient form in calculating the ground state energy as the following wave function which is suitable for fermions\cite{Jaminan80},
\begin{equation}
\Psi(r)=\frac{1}{\sqrt{A!}}{det\Psi_{i}(\vec{r_{i}})}\\
\end{equation}
Where the wave function of all nucleons $\Psi(r)$, and the wave function for i-nucleon $\Psi_{i}(r)$.
The wave function for nucleon $i$ depends on the oscillator parameter as,
\begin{equation}
 \Psi_{i}(\vec{r_{i}})= C_{i\alpha} F_{\alpha}(\vec{r_{i}})
\end{equation}
Where $C_{i\alpha}$ is the oscillator constant and $F_{\alpha}$ is the wave function of two components.
\begin{equation}
F_{\alpha} =\left|\begin{array}{c}
               \Phi_{\alpha} \\
               \chi_{\alpha}
             \end{array}\right\rangle
 \end{equation}
With the wave function for radial component $\Phi_{\alpha}$, and the spin component $\chi_{\alpha}$.
The principle of antisymmetry of the wave function was not completely explained by the Hartree method according to Slater and Fock independently. So the accurate picture in calculating the ground state energy is the Hartree-Fock approximation.
\begin{eqnarray}\label{r2}
\nonumber&&\sum_{i\alpha\beta}h_{i}C_{i\alpha}^{*}C_{i\beta}\left\langle F_{\alpha}|F_{\beta}\right\rangle=\\
\nonumber&&\sum_{i\alpha\beta}C_{i\alpha}^{*}C_{i\beta}\left\langle F_{\alpha}(r)|c\vec{\alpha_{i}}.\vec{p}+(\beta_{i}-I)m_{i}c^2|F_{\beta}\right\rangle \\
 \label{rabab2}&+&\sum_{i<j}\sum_{\alpha\gamma\beta\delta}C_{i\alpha}^{*}C_{i\beta}C_{j\gamma}^{*}C_{j\delta} \left\langle F_{\alpha}F_{\gamma}|(\frac{2}{m A}P_{ij}^{2}+V_{ij})|\widetilde{F_{\beta}F_{\delta}}\right\rangle
\end{eqnarray}
Where $C_{i\beta}$ is the occupation number or the oscillator number( For a system consisting of fermions, or particles with half-integral spin, the occupation numbers may take only two values; $0$ for empty states or $1$ for filled states). The two components of the wave functions have the following relation between them\cite{Bouyssy87,Long}.
\begin{equation}\label{r3}
\chi=\left(1-\frac{\varepsilon-v}{2Mc^2}\right)\frac{\vec{\sigma}.\vec{p}}{2mc}\phi
\end{equation}
Using the relation Eq (\ref{r3}), where $\varepsilon$ is external energy which equals zero, here we are deal with ground state and $c^3$ makes the value so small and can be neglected.
\begin{equation}\label{roby}
\chi\cong\frac{\vec{\sigma}.\vec{p}}{2mc}\phi
\end{equation}
\begin{eqnarray}
 \left\langle\Psi_{i}(r)|E|\Psi_{i}(r)\right\rangle   &=&\left\langle\Psi_{i}(r)|\hat{H_{1}}|\Psi_{i}(r)\right\rangle+\left\langle\Psi_{i}(r)|\hat{H_{2}}|\Psi_{i}(r)\right\rangle
 \end{eqnarray}
Differentiate Eq (\ref{rabab2}) with respect to $C_{i\alpha}$, the $\widetilde{F_{\beta}F_{\delta}}$ has a sign defines the exchange that happening between the two nucleons, hence substituting\\ $\sum_{j}C_{j\gamma}^{*}C_{j\delta}=1$ and $\langle F_{\alpha}|F_{\beta}\rangle=1$.
\begin{eqnarray}
\nonumber&&\sum_{i\alpha\beta}C_{i\beta}\left[\left\langle F_{\alpha}|\hat{H_{1}}|F_{\beta}\right\rangle+\sum_{i<j}\left\langle F_{\alpha}F_{\gamma}|\hat{H_{2}}|\widetilde{F_{\beta}F_{\delta}}\right\rangle-h_{i}\right]=\\
\nonumber&&\sum_{i\alpha\beta}C_{i\beta}[\left\langle F_{\alpha}|c\vec{\alpha_{i}}.\vec{p}+(\beta_{i}-1)m_{i}c^2|F_{\beta}\right\rangle\\
\nonumber&&+\sum_{i<j} \left\langle F_{\alpha}F_{\gamma}|(\frac{2}{m_{i}}P_{ij}^{2}+V_{ij})|\widetilde{F_{\beta}F_{\delta}}\right\rangle
-h_{i}]\\
\label{ano}&&= 0
\end{eqnarray}
Treating with the 1st part of Eq (\ref{ano}) give us the coming formula
\begin{equation}
H_{1}=\sum_{i\alpha\beta}C_{i\beta}\left\langle F_{\alpha}|c\vec{\alpha_{i}}.\vec{p}+(\beta_{i}-I)m_{i}c^2|F_{\beta}\right\rangle
\end{equation}
Taking into account Dirac matrices\cite{Green} with defining the wave functions in bracket as
 $|F_{\alpha}\rangle=|\begin{array}{c}
                                                    \phi_{\alpha} \\
                                                    \chi_{\alpha}
                                                  \end{array}\rangle
$ and $\langle F_{\alpha}|=\langle\begin{array}{cc}
                                                               \phi_{\alpha} & \chi_{\alpha}
                                                             \end{array}|
                                                  $ ,\\
$\vec{\alpha}=\left(
                        \begin{array}{cc}
                          0 & \vec{\sigma} \\
                          \vec{\sigma} & 0 \\
                        \end{array}
                      \right)$
, $\beta=\left(
   \begin{array}{cc}
     I & 0 \\
     0 & -I \\
   \end{array}
 \right)$
 where unit matrix $ I=\left(
\begin{array}{cc}
 1 & 0 \\
  0 & 1 \\
   \end{array}
   \right) $.\\
Substituting $\alpha, \beta$ in $H_{1}$ and have the result
   \begin{eqnarray}
\nonumber&& \langle F_{\alpha}|H_{1}|F_{\beta}\rangle \\
\nonumber&& =\left\langle F_{\alpha}|c\left(
                                                                    \begin{array}{cc}
                                                                      0 & \vec{\sigma} \\
                                                                      \vec{\sigma} & 0 \\
                                                                    \end{array}
                                                                  \right).\vec{p}+\left[\left(
   \begin{array}{cc}
     I & 0 \\
     0 & -I \\
   \end{array}
 \right)-
  \left(
\begin{array}{cc}
 1 & 0 \\
  0 & 1 \\
   \end{array}
   \right) \right]m_{i}c^2|F_{\beta}\right\rangle\\
\nonumber &&=\left\langle\phi_{\alpha}\left|\frac{(\vec{\sigma}.\vec{p})(\vec{\sigma}.\vec{p})}{2 m }\right|\phi_{\beta}\right\rangle + \left\langle\phi_{\alpha}\left|\frac{(\vec{\sigma}.\vec{p})(\vec{\sigma}.\vec{p})}{2 m }\right|\phi_{\beta}\right\rangle\\
\nonumber&&-\left\langle\phi_{\alpha}\left|\frac{(\vec{\sigma}.\vec{p})(\vec{\sigma}.\vec{p})}{ m }\right|\phi_{\beta}\right\rangle\\
&&= 0
\end{eqnarray}
The 1st term of kinetic energy tends to zero and this result has agreement with another calculations in \cite{Kample}
After the treatment of the kinetic terms are done, the residual Hamiltonian of the expectation value becomes,
\begin{eqnarray}
 H &=& H_{2} = \sum_{i<j}(\frac{2}{m_{i}}p_{ij}^{2}+V_{ij})
\end{eqnarray}
The popular form of the force between two nucleons is cleared according to meson exchanges. The potential form of one boson exchange $V_{ij}$ between two nucleons $(i,j)$ based on the degrees of freedom associated with three mesons, pseudoscalar, scalar, and vector mesons.
\begin{equation} \label{mohamed}
 V_{ij}(r)= V_{\pi}(r) + V_{\sigma}(r)+ V_{\omega}(r)+ V_{\rho}(r)
\end{equation}
The Dirac representation for functions of mesons  will be used and Dirac matrices corresponding to pauli spin matrices \cite{Miller,Anselm}, according to this representation we use $V_{\omega}(r)= V_{\rho}(r)$.
\begin{eqnarray} \label{mohamed2}
&& V_{ps}(r) =\gamma_{i}^{o}\gamma_{i}^{5}\gamma_{j}^{o}\gamma_{j}^{5}J_{ps}\\
&&V_{\sigma}(r)=-\gamma_{i}^{o}\gamma_{j}^{o}J_{\sigma}\\
&& V_{\omega}(r)=\gamma_{i}^{o}\gamma_{j}^{o}\vec{\gamma_{i}}^{\mu}\vec{\gamma_{j}}^{\mu}J_{\omega}\\
&&\vec{\gamma_{i}}^{\mu}\vec{\gamma_{j}}^{\mu}=[\gamma_{i}^{o}\gamma_{j}^{o}-\vec{\gamma_{i}}\vec{\gamma_{j}}]
 \end{eqnarray}
 Where
\begin{eqnarray} \label{mohamed3}
 \beta \equiv \gamma_{i}^{o}=\left(
                  \begin{array}{cc}
                    1 & 0 \\
                    0 & -1 \\
                  \end{array}
                \right)
 \; \vec{\gamma_{i}}=\left(
                                             \begin{array}{cc}
                                               0 & \vec{\sigma} \\
                                               -\vec{\sigma} & 0 \\
                                             \end{array}
                                           \right)\;
 \gamma^{5} = \imath\gamma^{o}\gamma^{1}\gamma^{2}\gamma^{3} =\left(
                          \begin{array}{cc}
                            0 & I \\
                            I & 0 \\
                          \end{array}
                        \right)
\end{eqnarray}
Substituting Eq. (\ref{mohamed2}) and Eq. (\ref{mohamed3}) into Eq. (\ref{mohamed}) to get the expectation value by three potentials $V_{\pi}$, $V_{\sigma}$ and $V_{\omega}$.
\begin{eqnarray}
\nonumber\langle F_{\alpha}F_{\gamma}| V_{ij}(r)|\widetilde{F_{\beta}F_{\delta}}\rangle
&=&\langle F_{\alpha}F_{\gamma}|V_{ps}|\widetilde{F_{\beta}F_{\delta}}\rangle+\langle F_{\alpha}F_{\gamma}|V_{s}|\widetilde{F_{\beta}F_{\delta}}\rangle+\langle F_{\alpha}F_{\gamma}|2 V_{v}|\widetilde{F_{\beta}F_{\delta}}\rangle\\
\nonumber&=&\langle F_{\alpha}F_{\gamma}|V_{\pi}|\widetilde{F_{\beta}F_{\delta}}\rangle+\langle F_{\alpha}F_{\gamma}|V_{\sigma}|\widetilde{F_{\beta}F_{\delta}}\rangle+\langle F_{\alpha}F_{\gamma}|2 V_{\omega}|\widetilde{F_{\beta}F_{\delta}}\rangle
\end{eqnarray}
\begin{eqnarray}
\nonumber \langle F_{\alpha}F_{\gamma}| V_{ij}(r)|\widetilde{F_{\beta}F_{\delta}}\rangle
&=& \langle\left(
                       \begin{array}{cc}
                         \phi_{\alpha} & \chi_{\alpha} \\
                       \end{array}
                     \right)
  |\langle\left(
            \begin{array}{cc}
              \phi_{\gamma} & \chi_{\gamma} \\
            \end{array}
          \right)
  |\left(\begin{array}{cc}
                1 & 0 \\
                0 & -1
                \end{array}\right)_{i}\left(
                \begin{array}{cc}
                0 & 1 \\
                 1 & 0 \\
                  \end{array}
                  \right)_{i}\\
                   \nonumber&& \left(
                  \begin{array}{cc}
                   1 & 0 \\
                    0 & -1 \\
                     \end{array}
                      \right)_{j}
                     \left(
                                                                                                                                                                              \begin{array}{cc}
                       0 & 1 \\
                       1 & 0 \\
                       \end{array}
                       \right)_{j} J_{\pi}
                        |\left(
                       \begin{array}{c}
                       \phi_{\beta} \\
                       \chi_{\beta} \\
                       \end{array}
                       \right)\rangle|
                       \left(
                       \begin{array}{c}
                       \phi_{\delta} \\
                       \chi_{\delta} \\
                       \end{array}
                       \right)\rangle\\
             \nonumber&+&\langle\left(
                       \begin{array}{cc}
                         \phi_{\alpha} & \chi_{\alpha} \\
                       \end{array}
                     \right)
  |\langle\left(
            \begin{array}{cc}
              \phi_{\gamma} & \chi_{\gamma} \\
            \end{array}
          \right)
  |\left(\begin{array}{cc}
                -1 & 0 \\
                0 & 1
                \end{array}\right)_{i}\left(
                \begin{array}{cc}
                1 & 0 \\
                 0 & -1 \\
                  \end{array}
                  \right)_{j}J_{\sigma} \\
                  \nonumber&&|\left(
                       \begin{array}{c}
                       \phi_{\beta} \\
                       \chi_{\beta} \\
                       \end{array}
                       \right)\rangle|
                       \left(
                       \begin{array}{c}
                       \phi_{\delta} \\
                       \chi_{\delta} \\
                       \end{array}
                       \right)
                       \rangle\\
\nonumber&+&2\langle
\left(\begin{array}{cc}
                         \phi_{\alpha} & \chi_{\alpha} \\
                       \end{array}
                     \right)
  |\langle
  \left(
            \begin{array}{cc}
              \phi_{\gamma} & \chi_{\gamma} \\
            \end{array}
          \right)
  |
  \left(\begin{array}{cc}
                1 & 0 \\
                0 & 1
                \end{array}\right)_{i}\left(
                  \begin{array}{cc}
                   1 & 0 \\
                    0 & 1 \\
                     \end{array}
                      \right)_{j} J_{\omega}\\
                      \nonumber&&|
                      \left(
                       \begin{array}{c}
                       \phi_{\beta} \\
                       \chi_{\beta} \\
                       \end{array}
                       \right)
                       \rangle|
                       \left(
                       \begin{array}{c}
                       \phi_{\delta} \\
                       \chi_{\delta} \\
                       \end{array}
                       \right)
                       \rangle \\
\nonumber&-&2 \langle\left(
                       \begin{array}{cc}
                         \phi_{\alpha} & \chi_{\alpha} \\
                       \end{array}
                     \right)
  |\langle\left(
            \begin{array}{cc}
              \phi_{\gamma} & \chi_{\gamma} \\
            \end{array}
          \right)
  |\left(\begin{array}{cc}
                0 & \vec{\sigma} \\
                \vec{\sigma} & 0
                \end{array}\right)_{i}\left(\begin{array}{cc}
                0 &\vec{ \sigma} \\
                \vec{\sigma} & 0
                \end{array}\right)_{j}J_{\omega}\\
                &&|\left(
                       \begin{array}{c}
                       \chi_{\beta} \\
                       -\phi_{\beta} \\
                       \end{array}
                       \right)\rangle
                       |
                       \left(
                       \begin{array}{c}
                       \phi_{\delta} \\
                       \chi_{\delta} \\
                       \end{array}
                       \right)\rangle
\end{eqnarray}
According to the relation between $\phi$ , $\chi$ in Eq.(\ref{roby}), one obtain
\begin{eqnarray}
\nonumber &&\langle F_{\alpha}F_{\gamma}| V_{ij}(r)|\widetilde{F_{\beta}F_{\delta}}\rangle= \\
\nonumber&&\langle\phi_{\alpha}\phi_{\gamma}|
                         \frac{1}{4 m^2 c^2}[J_{\pi}(\vec{\sigma_{j}}.\vec{p_{j}}) - (\vec{\sigma_{j}}.\vec{p_{j}})J_{\pi}(\vec{\sigma_{i}}.\vec{p_{i}}) - (\vec{\sigma_{i}}.\vec{p_{i}})J_{\pi}(\vec{\sigma_{j}}.\vec{p_{j}})\\
                \nonumber&&     + (\vec{\sigma_{i}}\vec{.p_{i}})(\vec{\sigma_{j}}.\vec{p_{j}})J_{\pi}]|
                        -J_{\sigma}+2 J_{\omega}+\frac{1}{4 m^{2}c^{2}}              [(\vec{\sigma_{i}}.\vec{p_{i}})J_{\sigma}(\vec{\sigma_{i}}.\vec{p_{i}})\\
                         \nonumber&&+(\vec{\sigma_{j}}.\vec{p_{j}})J_{\sigma}(\vec{\sigma_{j}}.\vec{p_{j}})
+2 (\vec{\vec{\sigma_{i}}}.\vec{p_{i}})J_{\omega}(\vec{\sigma_{i}}.\vec{p_{i}})
 +(\vec{\sigma_{j}}.\vec{p_{j}})J_{\omega}(\vec{\sigma_{j}}.\vec{p_{j}})\\
\nonumber&&-2 J_{\omega}(\vec{\sigma_{i}}.\vec{\sigma_{j}})(\vec{\sigma_{j}}.\vec{p_{j}})(\vec{\sigma_{i}}.\vec{p_{i}})
-2(\vec{\sigma_{j}}.\vec{p_{j}})J_{\omega}(\vec{\sigma_{i}}.\vec{\sigma_{j}})(\vec{\sigma_{i}}.\vec{p_{i}})
-2(\vec{\sigma_{i}}.\vec{p_{i}})J_{\omega} (\vec{\sigma_{i}}.\vec{\sigma_{j}})(\vec{\sigma_{j}}.\vec{p_{j}})\\
\label{13}&&-2(\vec{\sigma_{i}}.\vec{p_{i}})(\vec{\sigma_{j}}.\vec{p_{j}})J_{\omega}(\vec{\sigma_{i}}.\vec{\sigma_{j}})]|\widetilde{\phi_{\beta}\phi_{\delta}}\rangle
 \end{eqnarray}
Defining the momentum for each nucleon (i, j) $\vec{p_{i}}=\vec{p_{r}}+\frac{1}{2}\vec{p_{R}}$, $\vec{p_{j}}=-\vec{p_{r}}+\frac{1}{2}\vec{p_{R}}$ \cite{Neff,Gross}. Substituting those relations into Eq. (\ref{13}),
where $\vec{p_{r}}=\vec{p}$ and $(\vec{\sigma_{i}}.\vec{p_{R}})(\vec{\sigma_{i}}.\vec{p_{R}})=p_{R}^{2}$, we obtain
\begin{eqnarray}
\nonumber V_{ij}(r) &=&-J_{\sigma}+2 J_{\omega}+\frac{1}{4 m^{2}c^{2}}[(\vec{\sigma_{i}}.\vec{p})J_{\sigma}(\vec{\sigma_{i}}.\vec{p})
              + (\vec{\sigma_{j}}.\vec{p})J_{\sigma}(\vec{\sigma_{j}}.\vec{p})\\
              \nonumber&+&2 (\vec{\sigma_{i}}.\vec{p})J_{\omega}
              (\vec{\sigma_{i}}.\vec{p})+2 (\vec{\sigma_{j}}.\vec{p})J_{\omega}(\vec{\sigma_{j}}.\vec{p})+2 J_{\omega}(\vec{\sigma_{i}}.\vec{\sigma_{j}})(\vec{\sigma_{j}}.\vec{p})(\vec{\sigma_{i}}.\vec{p})\\
       \nonumber&&+2(\vec{\sigma_{i}}.\vec{p})J_{\omega}(\vec{\sigma_{i}}.\vec{\sigma_{j}})(\vec{\sigma_{j}}.\vec{p})+2 (\vec{\sigma_{j}}.\vec{p})
       J_{\omega}(\vec{\sigma_{i}}.\vec{\sigma_{j}})(\vec{\sigma_{i}}.\vec{p})+2(\vec{\sigma_{i}}.\vec{p})
              (\vec{\sigma_{j}}.\vec{p})\\
             \nonumber&& J_{\omega}(\vec{\sigma_{i}}.\vec{\sigma_{j}})+2 J_{\omega}(\vec{\sigma_{i}}.\vec{\sigma_{j}})(\vec{\sigma_{j}}.\vec{p})(\vec{\sigma_{i}}.\vec{p}_{R})-2  J_{\omega}(\vec{\sigma_{i}}.\vec{\sigma_{j}})(\vec{\sigma_{j}}.\vec{p_{R}})(\vec{\sigma_{i}}.\vec{p})\\
              \nonumber&&+2 (\vec{\sigma_{j}}.\vec{p})(\vec{\sigma_{i}}.\vec{p_{R}})J_{\omega}(\vec{\sigma_{i}}.\vec{\sigma_{j}})
              -2 (\vec{\sigma_{i}}.\vec{p})(\vec{\sigma_{j}}.\vec{p_{R}})
              J_{\omega}(\vec{\sigma_{i}}.\vec{\sigma_{j}})]\\
 \nonumber&&+ \frac{1}{8 m^{2}c^{2}}[(\vec{\sigma_{i}}.\vec{p})(\vec{\sigma_{i}}.\vec{p_{R}})J_{\sigma}+J_{\sigma}(\vec{\sigma_{i}}.\vec{p_{R}})(\vec{\sigma_{i}}.\vec{p})
              - (\vec{\sigma_{j}}.\vec{p})(\vec{\sigma_{j}}.\vec{p_{R}})J_{\sigma}\\
              \nonumber&&-J_{\sigma}(\vec{\sigma_{j}}.\vec{p_{R}})
              (\vec{\sigma_{j}}.\vec{p})+2 (\vec{\sigma_{i}}.\vec{p})(\vec{\sigma_{i}}.\vec{p_{R}})J_{\omega}+ 2J_{\omega}(\vec{\sigma_{i}}.\vec{p_{R}})(\vec{\sigma_{i}}.\vec{p})\\
              \nonumber&&-2 (\vec{\sigma_{j}}.\vec{p})(\vec{\sigma_{j}}.\vec{p_{R}})J_{\omega}
              -2J_{\omega}(\vec{\sigma_{j}}.\vec{p_{R}})(\vec{\sigma_{j}}.\vec{p})+\vec{p_{R}}^{2}J_{\sigma}
              +2 \vec{p_{R}}^{2}J_{\omega}\\
         \nonumber     &&-2 J_{\omega}(\vec{\sigma_{i}}.\vec{\sigma_{j}})(\vec{\sigma_{j}}.\vec{p_{R}})(\vec{\sigma_{i}}.\vec{p_{R}})]\\
          \nonumber    &&+
              \frac{1}{4 m^2 c^2}[-J_{\pi}(\vec{\sigma_{j}}.\vec{p})(\vec{\sigma_{i}}.\vec{p}) + (\vec{\sigma_{j}}.\vec{p})J_{\pi}(\vec{\sigma_{i}}.\vec{p}) + (\vec{\sigma_{i}}.\vec{p})J_{\pi}(\vec{\sigma_{j}}.\vec{p})\\
              &&- (\vec{\sigma_{i}}.\vec{p})(\vec{\sigma_{j}}.\vec{p})J_{\pi}]
\end{eqnarray}
 We will apply some important relations\cite{Raynal}
\begin{enumerate}
\item $(\vec{\sigma_{1}}.\vec{A})(\vec{\sigma_{1}}.\vec{B})=A.B+\imath\vec{\sigma_{1}}(A\times B)$
\item $(\vec{\sigma_{1}}.\vec{A})^{2}=A^{2}$
\item$(\vec{\sigma_{1}}.\vec{A})(\vec{\sigma_{2}}.\vec{A})=\frac{2}{\hbar^{2}}(S.A)^{2}-A^{2}$
\item $(\vec{\sigma}.\vec{A})F(r)(\vec{\sigma}.\vec{A})=F(r)A^{2}-\imath\hbar\{\bigtriangledown F(r).A+\imath\vec{\sigma}[(\bigtriangledown F(r))\times A]\}$
\end{enumerate}
Including these relations in potential equation.
 we substitute every term by using the relation of angular momentum $\vec{L}=\vec{r}\times \vec{p}$,
 $\vec{\sigma}=\frac{2\vec{S}}{\hbar}$ (where $\vec{S}$ is the total spin operator), $\vec{p}=-\imath\hbar\nabla$, and $\nabla J_{\sigma}=\frac{1}{r}(\frac{d J_{\sigma}}{d r})r$. According to the previous relations, and where $\vec{\sigma_{j}}^{2}=\vec{\sigma_{x}}^{2}+\vec{\sigma_{y}}^{2}+\vec{\sigma_{z}}^{2}=1$ as triplet case for two nucleons
 \begin{eqnarray}
\nonumber(\vec{\sigma_{j}}.\vec{p})J_{\omega}(r)(\vec{\sigma_{j}}\vec{\sigma_{j}})(\vec{\sigma_{j}}.\vec{p})&=&(\vec{\sigma_{j}}.\vec{p})J_{\omega}(r)
\vec{\sigma_{j}}^{2}(\vec{\sigma_{i}}.\vec{p})\\
\nonumber     &=&-3J_{\omega}(r)p^{2}+3\hbar^{2}\left\{\frac{d J_{\omega}}{d r}\frac{d}{d r}\right\}\\
     &-&\frac{6}{r}\frac{d J_{\omega}}{d r}[\vec{S_{j}}.\vec{L}]
\end{eqnarray}
 \begin{eqnarray}
\nonumber   (\vec{\sigma_{i}}.\vec{p})J_{\omega}(r)(\vec{\sigma_{i}}\vec{\sigma_{j}})(\vec{\sigma_{j}}.\vec{p})&=& -3 J_{\omega}(r)p^{2}+3\hbar^{2}\left\{\frac{d J_{\omega}}{d r}\frac{d}{d r}\right\}\\
   &-&\frac{6}{r}\frac{d J_{\omega}}{d r}[\vec{S_{i}}.\vec{L}]
 \end{eqnarray}
With total spin operator $\vec{S}$  and the meson function $J(r)$, using\cite{Varshalovich}\\
$(\vec{S}.\vec{p})^{2}=(\vec{S}.\hat{n})^{2}p^{2}$, $(\vec{\sigma_{i}}.\vec{\sigma_{j}})=\frac{2}{\hbar^{2}}S^{2}-3$ and $\vec{S}.\vec{L}=\frac{\hbar^{2}}{2}[J(J+1)-L(L+1)-S(S+1)]$ .
Quantum mechanics have a magnificent tool, this tool is the harmonic oscillator which is capable  of being solved in closed form, it has generally useful approximations and exact solutions of different problems\cite{Kirson}. It solves the differential equations in quantum mechanics.
We have the energy of Harmonic Oscillator$(\hbar\omega(2n+l+3/2))$ which equals the kinetic energy$(\frac{p^{2}}{2 m})$ added to the potential energy$((1/2)m\omega^{2}x^{2})$ to simplify the solution and get the result. It is slitted in relative harmonic oscillator energy
$\hbar\omega(2n+l+\frac{3}{2}) = \frac{p^{2}}{2 \mu}+\frac{1}{2}\mu\omega^{2}r^{2}$\cite{Gartenhaus,Gad}, with $\omega$ that is the angular frequency, and center of mass contribution in harmonic oscillator energy $\hbar\omega(2N+L+\frac{3}{2})= \frac{p^{2}}{2 M}+\frac{1}{2}M\omega^{2}R^{2}$.We suppose the nucleons have average masses $\frac{m_{n}+m_{p}}{2}$, so the relative mass $\mu=\frac{m_{1}m_{2}}{m_{1}+m_{2}}=\frac{m}{2}$, and center mass $M=m_{1}+m_{2}=2m$.
\begin{eqnarray}
\nonumber V_{ij}(r) &=& -J_{\sigma}+2J_{\omega}+\frac{1}{8 \mu^{2} c^{2}}[-\hbar^{2}\{\frac{d J_{\sigma}}{d r}\frac{d}{d r}\}+\frac{1}{r}\frac{d J_{\sigma}}{d r}[\frac{\hbar^{2}}{2}[J(J+1)\\
\nonumber&&-L(L+1)-S(S+1)]]+ 4\hbar^{2}\left\{\frac{d J_{\omega}}{d r}\frac{d}{d r}\right\}-2\frac{2}{r}\frac{d J_{\omega}}{d r}
             [\frac{\hbar^{2}}{2}[J(J+1)\\
             \nonumber&&-L(L+1)-S(S+1)]]+\frac{1}{4 \mu c^{2}}[J_{\sigma}(r)(\frac{\vec{p}^{2}}{2\mu})-
              2J_{\omega}(r)(\frac{\vec{p}^{2}}{2\mu})+ 4J_{\omega}\\
              \nonumber&&(2\vec{S}(\vec{S}+1)-3)(\frac{2}{\hbar^{2}}(\vec{S}.\hat{n})^{2}-1)(\frac{\vec{p}^{2}}{2\mu})+4(\frac{2}{\hbar^{2}}(\vec{S}.\hat{n})^{2}-1)
              (\frac{\vec{p}^{2}}{2\mu})J_{\omega}\\
              \nonumber&&(2\vec{S}(\vec{S}+1)-3)]+2\frac{1}{Mc^2}[-2(2\vec{S}(\vec{S}+1)-3)J_{\omega}(r)(\frac{2}{\hbar^{2}}(\vec{S}.\hat{n})^{2}-1)\\
              \nonumber&&(\frac{\vec{p_{R}}^{2}}{2M})
 + (\frac{p_{R}^{2}}{2M})J_{\sigma}+(\frac{p_{R}^{2}}{2M})
              J_{\omega}]++\frac{1}{4 m^2 c^2}[-J_{\pi}(2(\vec{S}.\hat{n})^{2}p^{2}+J_{\pi}p^{2}\\
\nonumber&&   -2\hbar^{2}(2\vec{S}(\vec{S}+1)-3)\frac{d J_{\pi}}{d r}\frac{d}{d r}
    -2(\vec{S}.\hat{n})^{2}p^{2}J_{\pi}+p^{2}J_{\pi} ]
\end{eqnarray}
The wave functions of the two nucleons Eq (\ref{13})should be treated as following .
\begin{eqnarray}
\nonumber&&\langle \phi_{\alpha}(r_{i})\phi_{\gamma}(r_{j})|=   \\
\nonumber&&\sum_{m_{l_{\alpha}}m_{s_{\alpha}}}\sum_{m_{l_{\gamma}}m_{s_{\gamma}}}
(l_{\alpha} s_{\alpha} m_{l_{\alpha}} m_{s_{\alpha}}|j_{\alpha} M_{\alpha})(l_{\gamma} s_{\gamma} m_{l_{\gamma}} m_{s_{\gamma}}|j_{\gamma} M_{\gamma})\\
&&\langle\phi_{n_{\alpha}l_{\alpha}m_{l_{\alpha}}}(r_{i})\phi_{n_{\gamma}l_{\gamma}m_{l_{\gamma}}}(r_{j})|
 \langle\chi_{m_{s_{\alpha}}}^{1/2}
 \chi_{m_{s_{\gamma}}}^{1/2}|\langle\hat{P}_{T_{\alpha}}\hat{P}_{T_{\gamma}}|
\end{eqnarray}
See \textbf{appendix B} to have the final formula.
\begin{eqnarray}
\nonumber&&\langle \phi_{\alpha}(r_{i})\phi_{\gamma}(r_{j})|=\\
\nonumber&&\sum_{m_{l_{\alpha}}m_{s_{\alpha}}}\sum_{m_{l_{\gamma}}m_{s_{\gamma}}}\sum_{JM}
    \sum_{\lambda\mu}\sum_{n l N L}\sum_{m M}\sum_{s m_{s}}\sum_{T}(l_{\alpha} s_{\alpha} m_{l_{\alpha}} m_{s_{\alpha}}|
                                                      j_{\alpha} M_{\alpha})(l_{\gamma} s_{\gamma} m_{l_{\gamma}} m_{s_{\gamma}}|j_{\gamma} M_{\gamma})\\
  \nonumber        &&(l_{\alpha}l_{\gamma}m_{l_{\alpha}}m_{l_{\gamma}}|\lambda\mu)
\langle n_{\alpha}l_{\alpha}n_{\gamma}l_{\gamma}|NL nl\rangle(l S m_{l} m_{S}|J M)(L l M m|\lambda\mu)\\
\nonumber&& (s_{\alpha}s_{\gamma}m_{s_{\alpha}}m_{s_{\gamma}}|S M_{s})
 (\chi_{\alpha}\chi_{\gamma}T_{\alpha}T_{\gamma}|T M_{T})\langle\phi_{NLM}(R)\phi_{nlm}(r)|\\
 && \langle\chi_{m_{s_{\alpha}}}^{1/2}(i,j)|\langle\hat{P}_{T}(i,j)|
\end{eqnarray}
The bracket $\langle n_{\alpha}l_{\alpha}n_{\gamma}l_{\gamma}|NL nl\rangle$ represents the Talmi Moshinsky bracket. The same treatment for the ket part $| \phi_{\beta}(r_{i})\phi_{\delta}(r_{j})\rangle$ to have
\begin{eqnarray}
\nonumber&&| \phi_{\beta}(r_{i})\phi_{\delta}(r_{j})\rangle=\\
\nonumber&&\sum_{m_{l_{\beta}}m_{s_{\beta}}}\sum_{m_{l_{\delta}}m_{s_{\delta}}}\sum_{JM}
    \sum_{\lambda\mu}\sum_{n l N L}\sum_{m M}\sum_{s m_{s}}\sum_{T}(l_{\beta} s_{\beta} m_{l_{\beta}} m_{s_{\beta}}|
                                                      j_{\beta} M_{\beta})(l_{\delta} s_{\delta} m_{l_{\delta}} m_{s_{\delta}}|j_{\delta} M_{\delta})\\
     \nonumber                                                 &&(l_{\beta}l_{\delta}m_{l_{\beta}}m_{l_{\delta}}|\lambda\mu)
  \langle n_{\beta}l_{\beta}n_{\delta}l_{\delta}|NL nl\rangle(l S m_{l} m_{S}|J M)(L l M m|\lambda\mu)\\
\nonumber&&(s_{\beta}s_{\delta}m_{s_{\beta}}m_{s_{\delta}}|S M_{s})
 (\chi_{\beta}\chi_{\delta}T_{\beta}T_{\delta}|T M_{T})|\phi_{NLM}(R)\phi_{nlm}(r)\rangle \\
 &&|\chi_{m_{s_{\beta}}}^{1/2}(i,j)|\hat{P}_{T}(i,j)\rangle
\end{eqnarray}
we have the wave function $\phi_{NLM}(R)=R_{NLM}(R)Y_{NLM}(\vartheta,\varphi)$ as radial part $(R)$ and angular part $(Y)$ for center of mass coordinates, the wave function $\phi_{nlm}(r)=R_{nlm}(r)Y_{nlm}(\vartheta,\varphi)$ as radial part $(R)$, and angular part $(Y)$ for relative coordinates.
The two-nucleons interaction formula through the exchange of four mesons where $\vec{p_{ij}}=\vec{p}$ and $A$ is the mass number of the required nuclei. We define
 the bracket $\langle\chi_{m_{s}}^{s}(i,j)|\chi_{m_{s}}^{s}(i,j)\rangle=1$, $\langle\hat{P}_{T}(i,j)|\hat{P}_{T}(i,j)\rangle =1$ and\\
$\langle Y_{N LM} Y_{nlm}|Y_{N LM} Y_{nlm}\rangle=1$ as the terms of equation depend on $(r)$.\\
We have the formula of radial wave function which involves the length parameter $b=\sqrt{\frac{\hbar}{m\omega}}$ with angular frequency $\omega$\\ and the associated Lageurre polynomial $L_{n}^{l+\frac{1}{2}}$.
\begin{equation}\label{r6}
R_{nl}=\left[\frac{2n!}{\Gamma(n+l+\frac{3}{2})}\right]^{\frac{1}{2}}\left(\frac{1}{b}\right)^{\frac{3}{2}}\left(\frac{r}{b}\right)^{l}
    exp\left(\frac{-1}{2}\left(\frac{r}{b}\right)^{2}\right)L_{n}^{l+\frac{1}{2}}\left(\frac{r}{b}\right)^{2}
\end{equation}
The differentiation of Radial function equals
\begin{equation}\label{r5}
\frac{d}{dr}R_{nl}(r)=\frac{l}{r}R_{n,l}-\frac{r}{b^{2}}R_{n,l}-\frac{2r\sqrt{\acute{n}}}{b^{2}}R_{n-1,l+1}
\end{equation}
Defining the operator $\vec{S}.\hat{n}$ \cite{Varshalovich}as
\begin{eqnarray}\label{r7}
\nonumber&& (S.\hat{n})Y_{JM}^{LS}(\vartheta,\varphi)=\\
\nonumber&&\frac{-\hbar}{2}[(\frac{(J+L+S+2)(J+L+S+1)(J-L+S)(-J+L+S+1)}{(2L+1)(2L+3)})^{\frac{1}{2}}\\
\nonumber&&Y_{JM}^{L+S}(\vartheta,\varphi)\\
\nonumber&&+(\frac{(J+L+S+1)(J+L-S)(J-L+S+1)(-J+L+S)}{(2L-1)(2L+1)})^{\frac{1}{2}}\\
&&Y_{JM}^{L-S}(\vartheta,\varphi)]
\end{eqnarray}
 The meson degrees of freedom have some static functions $J_{k}$ for description, but here we choose $GY$ and $SPED$ for meson $k$ and $(k= \pi,\sigma,\omega,\rho)$.
 \begin{equation}\label{robyy}
    (J_{k})_{GY}=g_{k}\hbar c \left(\frac{\exp(-\mu_{k}r)}{r}-\frac{\exp(-\lambda_{k}r)}{r}\left(1+\frac{\lambda_{}^{2}-\mu_{k}^{2}}{2\lambda_{k}}r\right)\right)
\end{equation}
Where the meson-nucleon coupling constant $g_{k}^{2}$, the cut off $\lambda_{k}$ and the mass of the meson is associated with $\mu_{k}=\frac{mc}{\hbar}$. The second function has the form \cite{Jaminan80},
\begin{equation}\label{robb}
    (J_{k})_{SPED}=g_{k}\hbar\left(\frac{\lambda_{k}^{2}}{\lambda_{k}^{2}-\mu_{k}^{2}}\right)\left(\frac{\exp(-\mu_{k}r)}{r}-\frac{\exp(-\lambda_{k}r)}{r}\right)
\end{equation}

 \section{Theoretical analysis of $CDP$}
There is an explicit spin dependence for the instanton interaction unlike one-gluon exchange, it contains a projector on spin $S=0$ states. The distribution of this interaction represented with $\delta (\vec{r})$ replaced by a Gaussian function with range $\Lambda$.
\begin{equation}\label{distribution}
\delta(r)= \frac{1}{\Lambda^{3}}\frac{1}{\pi^{\frac{3}{2}}}\exp(\frac{-r^{2}}{\Lambda^{2}})
\end{equation}
 with $\Lambda$ is the range of the pairing force($QCD$ scale parameter).
\begin{equation}\label{instanton potential}
    V_{I}(r) = 8\left(
                  \begin{array}{cc}
                    g & \sqrt{2}\grave{g} \\
                    \sqrt{2}\grave{g} & 0 \\
                  \end{array}
                \right) \delta (r)
\end{equation}
Where $g, \acute{g}$ are two dimensioned coupling constants according to quark flavors.This equation is under condition of $(l=s=0, I=0)$, where $l, s, I$ denotes angular momentum, spin and isotopic spin quantum numbers respectively for $n\bar{n}$ pair, and the form of instanton contributions represents as \cite{semay95}.
The pairing force depends on the value of the parameters $g$ and $\acute{g}$, if we set $g$ for strange flavor with symbol $s$ and $\grave{g}$ for non-strange flavor with symbol $n$. The Hamiltonian contains diagonal parts in the isoscalar space ($|n\bar{n}\rangle$,$|s\bar{s}\rangle$).
 \begin{equation}\label{hamiltonian instfina}
  H=\left(
      \begin{array}{cc}
        \sum^{3}_{i=1}\sqrt{\vec{p_{i}}^{2}+m_{i}^{2}} & 0 \\
        0 & \sum^{3}_{i=1}\sqrt{\vec{p_{i}}^{2}+m_{i}^{2}}\\
      \end{array}
    \right)+\frac{1}{2}[\frac{-k}{r}+ar+C]\left(
                                            \begin{array}{cc}
                                              1 & 0 \\
                                              0 & 1 \\
                                            \end{array}
                                          \right)+V_{I}(\vec{r})
\end{equation}
Having the coupling constant as,
 \begin{equation}\label{coupling}
    \grave{g}=\frac{3}{8}g_{eff}(n)
 \end{equation}
The parameter $g_{eff}$ denotes the strength and is defined as \cite{resag},
\begin{equation}\label{effec constant}
  g_{eff}= (\frac{4}{3}\pi^{2})^{2}\int_{0}^{\rho_{c}}d\rho d_{o}(\rho) \rho^{2}\times (m_{i}^{o}-\rho^{2} c_{i})
\end{equation}
Where $d_{o}(\rho)$ is a function instanton density of the instanton size $\rho$,  For three colors and three flavors this quantity is given in \cite{resag}, $m_{i}^{o}$ is the current mass of flavor $i$ and the quark condensate for this flavor is $c_{i}=(\frac{2}{3})\pi^{2}\langle \bar{q_{i}} q_{i}\rangle$, $\langle \bar{q_{i}} q_{i}\rangle$ (non-vanishing expectation values). The integration over $\rho_{c}$ which is the maximum size of the instanton.
\begin{equation*}
  d_{o}(\rho)= (3.63\times 10^{3})(\frac{8 \pi^{2}}{g^{2}(\rho)})^{6} \exp(\frac{8\pi^{2}}{g^{2}(\rho)})
\end{equation*}
Where
\begin{equation*}
(\frac{8 \pi^{2}}{g^{2}(\rho)})= 9 \ln(\frac{1}{\Lambda \rho})-\frac{32}{9}\ln(\ln(\frac{1}{\Lambda \rho}))
\end{equation*}
 The constituent masses are the re-normalization of quarks' masses which demonstrate the contribution of the constituent masses \cite{resag}.
\begin{equation}\label{constituent}
  m_{n}=m_{n}^{o}+\Delta m_{n}+\delta_{n}
\end{equation}
$m_{n}^{o}$ is the current mass of non-strange quark, $\Delta m_{n}$ is the contribution of constituent mass \cite{brausemay} with free parameter $\delta_{n}$ added to the running masses. The contribution of the constituent masses has the following formula.

\begin{equation*}
  \Delta m_{n}=\frac{3}{4} \pi^{2}\int_{0}^{\rho_{c}}d\rho d_{o}(\rho) \rho^{2} (m_{n}^{o}-\rho^{2} c_{n})(m_{s}^{o}-\rho^{2} c_{s})
\end{equation*}
It is important to replace the  dimensional instanton size \cite{cfb}
as $x=\Lambda \rho$ with a dimensionless quantity with using the definition of $d_{o}(\rho)$,
\begin{equation*}
  \alpha_{n}(x_{c})=\int_{0}^{x_{c}}dx [9 \ln(\frac{1}{x})-\frac{32}{9}\ln(\ln(\frac{1}{x}))]^{6}x^{n}[\ln(\frac{1}{x})]^{-\frac{32}{9}}
\end{equation*}
This dimensionless integration should still have small value of $\ln\ln$-term. It is involved in the parameters $\acute{g}, \Delta m_{n}$.
\begin{equation}\label{coupling 2}
  \grave{g}=\frac{\delta \pi^{2}}{2 \Lambda^{3}}[m_{n}^{o}\alpha_{11}(x_{c})-\frac{c_{n}}{\Lambda^{2}}\alpha_{13}(x_{c})]
\end{equation}
\begin{equation}\label{contribution2}
  \Delta m_{n}=\frac{\delta}{\Lambda}[ m_{n}^{o} m_{s}^{o}\alpha_{9}(x_{c})-\frac{c_{n} m_{s}^{o}+c_{s}m_{n}^{o}}{\Lambda^{2}}\alpha_{11}(x_{c})+\frac{c_{n}c_{s}}{\Lambda^{4}}\alpha_{13}(x_{c})]
\end{equation}
The functions $\alpha_{9}(x_{c}), \alpha_{11}(x_{c}), \alpha_{13}(x_{c})$ given in \cite{cfb}, the $m_{s}^{o}$ is the constituent mass of strange flavor and also the $c_{s}$ is the quark condensate related to the strange flavor. It is supposed that the quark as an effective degrees of freedom is dressed by the gluon and quark-anti quark pair clouds(constituent masses) and it is natural to express the probability density of quark configuration as a Gaussian function around its average position.
\begin{equation*}
  \rho_{i}(r)=\frac{1}{(\gamma_{i}\sqrt{\pi})^{\frac{3}{2}}}\exp(\frac{-r^{2}}{\gamma_{i}^{2}})
\end{equation*}
Where $ \rho_{i}(r)$ is the probability density not the instanton size as previous with $\gamma_{i}$ the size parameter and it is dependent on the quark mass flavor($n$ for non-strange flavor and $s$ for strange flavor). The operator for the quark in positions $r_{1}$ and $r_{2}$ is replaced by effective one after double convolution of the bare operator with the density functions $\rho_{i}$ and $\rho_{j}$.  This can be performed by using the dressed expression $\tilde{O_{ij}}(r)$ of the bare operator $O_{ij}(r)$ which depends only on the relative distance $r_{ij}=r_{i}-r_{j}$ between quarks \cite{brausemay}.
\begin{equation}\label{dressed}
 \tilde{O_{ij}}=\int d r O_{ij}(\acute{r})\rho_{ij}(r_{ij}-\acute{r})
\end{equation}
The convolution procedure supposed to remain the center of mass fixed during it and that the $\rho_{ij}$ tends to a delta function at the limit of an infinitely large $\gamma_{ij}$ \cite{brac}.
\begin{equation*}
  \tilde{\delta}(r)=\frac{1}{(\gamma_{ij}\sqrt{\pi})^{3}}\exp(\frac{-r^{2}}{\gamma_{ij}^{2}})
\end{equation*}
This formula resembles the previous form of the probability density of Gaussian form, but with parameter $\gamma_{ij}$. The convolution (a function derived from two given functions by integration that expresses how the shape of one is modified by the other) of two Gaussian functions with size parameter $\gamma_{i}$ and $\gamma_{j}$ is also a Gaussian function. After convolution with the quark density, the Cornell dressed potential has the following form,
\begin{equation}\label{dressed potential}
  \tilde{V}_{C}(r)=-k \frac{erf(\frac{r}{\gamma_{ij}})}{r}+a r [\frac{\gamma_{ij}\exp(\frac{-r^{2}}{\gamma_{ij}^{2}})}{\sqrt{\pi} r}+(1+\frac{\gamma_{ij}^{2}}{2 r^{2}})erf(\frac{r}{\gamma_{ij}})]+C
\end{equation}
With the error function $erf$ and $\gamma_{ij}=\sqrt{\gamma_{i}^{2}+\gamma_{j}^{2}}$, where $\gamma_{i}=\frac{1}{(\gamma_{i}\sqrt{\pi})^{\frac{3}{2}}}\exp(\frac{-r^{2}}{\gamma_{i}^{2}})$ and $\gamma_{j}=\frac{1}{(\gamma_{j}\sqrt{\pi})^{\frac{3}{2}}}\exp(\frac{-r^{2}}{\gamma_{j}^{2}})$.

 \section{Results and discussion}
\par
In table \ref{tab:abcdef} represents the group of parameters used for $(\pi, \sigma,
\omega, \rho)$ mesons. The set of parameters are I,II that include mass of
meson, the coupling constant $(g)$ and the cut-off parameter $(\lambda)$.
\begin{table}[h]
\centering
\caption{The meson parameters for $OBEP$ for different sets \cite{abdo}.}
\begin{tabular}{ | l| l | l | l |  l | l | }
 \hline       &Ref           & meson        & mass$ MeV$     &coupling constant       & Cut off parameter         \\
              &               &              &               &       $g_{i}/4\pi$                        &$\lambda$\;$MeV$           \\
 \hline &   &\;\;\; $\pi$    & \;\;\; $138.03$ &\;\;\;\;\;\;\;\;\; $14.9$     &\;\;\;\;\;\;\;\;\;\;\;\; $2000$      \\
              &              &\;\;\; $\sigma$ &\;\;\; $700$  &\;\;\;\;\;\;\;\;\; $16.07$      &\;\;\;\;\;\;\;\;\;\;\;\;\;\; $2000$      \\
        set I & \cite{abdo}  &\;\;\; $\omega$ &\;\;\; $782.6$&\;\;\;\;\;\;\;\;\; $28$      &\;\;\;\;\;\;\;\;\;\;\;\;\;\; $1300$      \\
              &              &\;\;\; $\rho$   &\;\;\; $769$  &\;\;\;\;\;\;\;\;\; $1.7$     &\;\;\;\;\;\;\;\;\;\;\;\;\;\; $1100$    \\
 \hline &  &\;\;\; $\pi$    &\;\;\; $138.03$ &\;\;\;\;\;\;\;\;\; $14.40$     &\;\;\;\;\;\;\;\;\;\;\;\;\;\; $1700$      \\
              &              &\;\;\; $\sigma$ &\;\;\; $710$  &\;\;\;\;\;\;\;\;\; $18.37$      &\;\;\;\;\;\;\;\;\;\;\;\;\;\; $2000$      \\
        set II & \cite{abdo}  &\;\;\; $\omega$ &\;\; $782.6$  &\;\;\;\;\;\;\;\;\; $24.50$     &\;\;\;\;\;\;\;\;\;\;\;\;\;\; $1850$ \\
              &              &\;\;\; $\rho$ &\;\; $769$  &\;\;\;\;\;\;\;\;\; $0.9$     &\;\;\;\;\;\;\;\;\;\;\;\;\;\; $1850$     \\   \hline
                         \end{tabular}

      \label{tab:abcdef}
\end{table}
\par
The parameters listed in table \ref{tab:abc} is related to the quark-quark potential through the $CDP$ which is added to the nucleon-nucleon potential in the hybrid model.
\begin{table}[h]
\centering
\caption{The quark parameters for the instanton induced interaction with the $CDP$ \cite{cfb}.}
\begin{tabular}{ | l| l | l |}
 \hline       Parameters           & Unit       & Values in baryon \\
 \hline       \;\;\; $a$    & \;\;\; $GeV^{2}$ &\;\;\; $0.16803$  \\
 \hline       \;\;\; $K$   &           &\;\;\; $0.79801$      \\
 \hline       \;\;\; $C$ &\;\;\; $GeV$ &\;\;\; $-0.96701$       \\
 \hline        \;\;\; $m_{n}$      &\;\;\; $GeV$   &\;\;\; $0.378$    \\
 \hline       \;\;\; $\gamma_{n}$  &\;\;\; $GeV^{-1}$    &\;\;\; $0.68101$ \\  \hline
  \end{tabular}

  \label{tab:abc}
\end{table}
\par
The two-body force is a simple model to reveal the hidden physics of the
atomic systems and also the nuclear systems. Our work boils down to simple
fact of constructing more realistic model that contains all possible degrees
of freedom in some light nuclei such as Deuteron $^{2}H_{1}$ and Helium
$^{4}H_{2}$. So, we include the interaction between two baryons which is
bounded in a hadron and each baryon contain three bounded quarks. The
nucleon-nucleon interaction is well introduced by the exchange of mesons with
the $OBE$ model. At long range of this interaction, it is supposed to be due
to the exchange of pion-meson(pseudoscalar meson) followed by the effect of
scalar meson $(\sigma)$ in attractive attitude at the medium range. The
attractive behavior have to face an opposite behavior to maintain the
stability of nuclei, so, the short range of this interaction is affected by a
repulsive behavior due to the exchange of vector mesons such as $(\omega,
\rho)$ and $QCD$ effects.
The potential is elaborated to calculate the ground state energies for the $^{2}H$, $^{4}He$
nuclei. We have examined the
$OBEP$ to calculate the ground state energy of
$H^{2}$ and $He^{4}$ nuclei using two static meson functions $(GY, SPED)$
with two sets of parameters listed in table \ref{tab:abcdef} which shows the
different sets of the used parameters and for different exchange mesons,
$(\sigma, \omega)$ mesons, $(\pi, \sigma, \omega)$ mesons and $(\pi, \sigma, \omega, \rho)$.We have mentioned that there is an effect of $QCD$ at the short range of
nucleon-nucleon interaction via three bounded quarks interacted between each
other. The so-called Funnel potential or Cornell potential which is the
simple and best model for the description of Charmonium system, but in our
hybrid model without coupling between mesons and quarks, it gives too high
values and the effect of our model is a destructive one. When we tried to
apply the idea of hybrid model with the aid of the instanton induced
interaction, it really gives us a transition probability for the interaction
of quark-quark interaction in small scale comparing with the confinement
scale. It is indeed similar to the tunneling effect with possibility of
treating the instanton interaction as a field configuration between quarks
and anti-quarks in the ground states. This interaction is also applied on the
light quarks not only the quark-antiquark. So, it is useful for us in our
model as the proton or neutron is a hadron of three light quarks. \\The
instanton interaction is included in the $CDP$, giving us
a small value ranged between $-0.15 MeV$ in case of $^{2}H$ and $-0.25MeV$ in
case of $^{4}He$ around the boundaries of the hadron. Our results shown in
table \ref{tab:abcdefghi} and table \ref{tab:abcdefghi2} with the effect of $CDP$ with parameters of table \ref{tab:abc} besides the exchange of
mesons through $OBEP$. Generally, the effect is encouraged and it improves
the ratios for the ground state energies of the Deuteron and Helium nuclei in
all cases with different parameters of meson degrees of freedom and different
functions $GY$ and $SPED$.\\
We have determined the ratio $Rat$, to ensure the accuracy between the
calculated results and the experimental data.
\begin{equation}\label{ads}
    Rat= \frac{E_{theor.} }{E_{exp.}}
\end{equation}
Where $E_{theor}$ is the calculated ground-state and $E_{exp}$ the
experimental one. We can also determine the binding energy per nucleon
$\frac{B.E}{A}$ for the studied nuclei as \cite{gad2011},
\begin{equation}\label{binding}
    \frac{B.E}{A}=-\frac{E_{g.s.}}{A}
\end{equation}
with the mass number $A$, and the total ground state energy $E_{g.s.}$.
The results of the $OBEP$ are listed in tables (\ref{tab:abcdefghijk}$-$\ref{tab:abcdefghijkl}) in
comparison with other theoretical and experimental data. The ratio between the present work and
experimental one is estimated for both cases, in other words by using the
potential extracted from $GY$ and $SPED$ functions.

\begin{table}[ht!]
\caption{The ground state energy of $^{2}H$ nucleus based on $OBEP$.}
\hspace{-2cm}
 \begin{tabular}
 { | l | l |l | l | l | l | l | l | l | l | l |}
 \hline {\small parameter} &   & {\small Present} & {\small Present} &others& exp. &Ratio  &Ratio&{\small ${B.E}/{A}$}&${B.E}/{A}$ \\
  &  meson & work& work  &   & \cite{garcona} & ${\small GY}$ & ${\small SPED}$&$GY$& $SPED$\\
  sets\cite{abdo}           &               &  $(GY)$        & $(SPED)$   &       & \cite{brner},   &      &      &    &   \\
                            &               &                 &            &       & \cite{kessler}                &     &      &    &    \\
\hline \;\;\; I           &($\sigma$, $\omega$) & $-2.916$&$-2.041$ & -2.215&           &1.311&0.918&1.458&1.0205\\
       \;\;\; II                &                       & $-3.486$&$-1.973$&\cite{haung} & $-2.224$ &1.567&0.887&1.743&0.987\\
       \;\;\; I            &($\pi$, $\sigma$,    & $-2.199$&$-2.248$& &  &0.989&1.011 &1.099&1.124\\
       \;\;\; II         &    $\omega$ )        & $-2.168$&$-2.204$ &    &        &0.975&0.991 &1.084&1.102\\
       \;\;\; I          & ($\pi$, $\sigma$, &$-2.127$&$-2.167$& &   &$0.9563$&$0.974$&1.063 &1.084 \\
\;\;\; II         &      $\omega$, $\rho$)     &$-1.877$&$-2.379$&       &   &$0.8438$&$1.069$ &0.938 &1.189 \\ \hline
   \end{tabular}
   \label{tab:abcdefghijk}
   \end{table}
At first, the $OBEP$ depended on the
cancelation of $\sigma$-meson and $\omega$-meson, the results is satisfied
for the ground state energies of $^{2}H$ nucleus as in table \ref{tab:abcdefghijk}
with two different sets of meson s' parameters, but here also tried to
include the $OBEP$ through the exchange of three mesons $(\pi, \sigma,
\omega)$ and four mesons $(\pi, \sigma, \omega, \rho)$, results are listed in
table \ref{tab:abcdefghijk} with two static functions for the meson; $GY$, $SPED$
functions. The preferable value of deuteron ground state is in case of using three mesons by using $SPED$ function for parameters $I$. It is noticed that the case of exchange three mesons gives closer value than the case of exchange of four mesons and demonstrating that the effect of $\pi$ meson as an attractive one to be clear than the effect of $\rho$ meson. This behavior is reasonable for light nuclei.\\
\begin{table}[ht!]
\caption{The ground state energy of $^{4}He$ nucleus through OBE.}
\hspace{-2cm}
\begin{tabular}
{ | l | l |l | l | l | l | l | l |l | l|l}
\hline {\small  Parameter}  &   &present& present &others& exp. & Ratio& Ratio&${E}/{A}$&${E}/{A}$\\
  &meson &work  &work  &\cite{kowalski}   & \cite{Tyren66} & $GY$ & $SPED$&$GY$& $SPED$\\
 sets\cite{abdo}  &            &$(GY)$  &   $(SPED)$   &   &                 &     &        &   &       \\
\hline \;\;\; I      & ($\sigma$, $\omega$)& $-22.372$   &$-20.238$& &              &1.0966&0.992&5.593&5.0595\\
       \;\;\; II             &                            &$-22.751$  &$-21.556$&-21.385 & -20.4&1.115&1.057&5.6877&5.389\\
       \;\;\; I     & ($\pi$, $\sigma$, &$-22.637$&$-20.375$&  &$\pm0.3$  &1.109&0.999&5.659&5.0937\\
       \;\;\; II &$\omega$ )  & $-21.871$&$-20.337$&   &        &1.072&0.997&5.4677&5.08425\\
       \;\;\; I     & ($\pi$, $\sigma$,  & $-19.655$&$-19.7388$&   &   &0.9497&0.9675&4.9137    & 4.9347\\
       \;\;\; II &  $\omega$,$\rho$)     & $-19.3744$&$-20.7917$&      &     &0.9497&1.0192&4.8436   &5.1979\\\hline
       \end{tabular}

       \label{tab:abcdefghijkl}
  \end{table}
\par
The $^{4}He$ nucleus has the same manner as the $^{2}H$ nucleus with preferable values ranged from $20.1-20.7$ and that is listed in table \ref{tab:abcdefghijkl}.\\
 The ratio is getting better result for going on more massive nuclei and encouraged for our potential. The calculation of binding energy per nucleon serves our idea of being the $OBEP$ with three and four mesons in case of $SPED$ function, and gives satisfied values for Deuteron and Helium nuclei comparing with the experimental one as it is for Deuteron $B.E/A=1.112$ and for Helium is $B.E/A\simeq5.1$.
\begin{table}[ht!]
\caption{The ground state energy of Deuteron with the hybrid model related to quark and meson degrees of freedom.}
\hspace{-2cm}
\begin{tabular}
{ | l | l |l | l | l | l | l |l|l| }
\hline {\small parameter}  & meson  & Hybrid & Hybrid & exp.\cite{garcona} & Ratio  & Ratio&${B.E}/{A}$&${B.E}/{A}$ \\
   & exchange & $(GY+$ & $(SPED+$ &\cite{brner,kessler} & $GY$ & $SPED$& $GY$ & $SPED$ \\
     sets\cite{abdo}&          &  $CDP)$ &  $CDP)$ &                              &     $+CD$         &    $+CD$ &   $+CD$         &    $+CD$       \\
\hline \;\; I    &\; ($\sigma$, $\omega$) & $-3.066$& $-2.191$ &          &1.3785&0.985&1.533    &1.096\\
\;\; II          &                       & $-2.496$& $-2.123$&   &1.122&0.955&1.248 &1.0615\\
\;\; I           &\; ($\pi$, $\sigma$,    & $-2.349$& $-2.398$& $-2.224$ &1.056&1.078&1.175 &1.199 \\
\;\; II         &   $\omega$)        & $-2.318$& $-2.354$ &        &1.042&1.058&1.159  &1.177\\
\;\; I          & \; ($\pi$, $\sigma$, & $-2.277$& $-2.317$&   &$1.024$&$1.042$&1.138&1.158 \\
\;\; II         &  $\omega$, $\rho$ )        &$-2.027$& $-2.529$&       &$0.911$&$1.1371$&1.033 &1.264\\ \hline
\end{tabular}

\label{tab:abcdefghi}
\end{table}

\begin{table}[ht!]
\caption{The ground state energy of Helium with the hybrid model.}
\hspace{-2cm}
\begin{tabular}
{ | l | l |l | l | l | l | l |l|l| }
\hline{\small parameter}  & meson  & Hybrid & Hybrid & exp. & Ratio  & Ratio&${B.E}/{A}$&${B.E}/{A}$ \\
    & exchange & $(GY+$ & $(SPED+$ & \cite{ Tyren66} & $GY$     & $SPED$ & $GY$     & $SPED$\\
     sets\cite{abdo}                  &          &  $CDP)$ &  $CDP)$ &                              &     $ +CD$        &  $+CD$   &     $ +CD$        &  $+CD$         \\
\hline \;\; I    & ($\sigma$, $\omega$) & $-22.622$& $-20.488$ &          & 1.1089&1.0043&5.655&5.122\\
\;\; II          &                       & $-23.001$& $-21.806$&   & 1.1275&1.0689&5.750&5.541\\
\;\; I           & ($\pi$, $\sigma$,   & $-22.887$& $-20.625$& $-20.4$ & 1.1219&1.0110&5.7217&5.156\\
\;\; II         &  $\omega$ )         & $-22.121$& $-20.587$ &  $\pm0.3$      & 1.084&1.0091&5.530&5.147\\
\;\; I          & ($\pi$, $\sigma$,  & $-19.905$& $-19.9888$&       & $0.9757$&$0.9798$&4.976&4.997\\
\;\; II         & $\omega$, $\rho$)         & $-19.6244$& $-21.0417$&       & $0.9619$&$1.03145$&4.906&5.260\\ \hline
\end{tabular}

\label{tab:abcdefghi2}
\end{table}
\par
Tables (\ref{tab:abcdefghi},\ref{tab:abcdefghi2}) have the effect of adding quark degrees of freedom to the meson degrees of freedom in a hybrid model for all previous cases, The values are reasonable and the best value of the hybrid model with the exchange of four mesons in case of parameters $II$ for $^{2}H$ nucleus when we apply the $GY$ function than others. The results are different for $^{4}He$ nucleus, we have the value of $SPED$  function with the exchange of two, three mesons in the hybrid model with the parameters $I$ to be the preferred one. It is obvious from table \ref{tab:abcdefghi} and table \ref{tab:abcdefghi2} the ground energy is close to the data in case of SPED function for set $I$ and set $II$ in comparison with the experimental data. The $^{4}He$ nucleus has little different manner, the theoretical values of the hybrid model are more cleared than in $^{2}H$ nucleus. It is noticed that the quark-quark interaction improves the values with $GY$ function. We concluded that the used
model is well-defined and  compatible with the data and even than other models see \cite{Wiringa,Iyad18}.
 \section{Conclusion}
In the framework of quasi relativistic formulation, the meson exchange
potential helps in obtaining a potential with few number of parameters to
calculate the ground state for the light nuclei Deuteron and Helium using two
$(\sigma,\omega)$, three $(\pi,\sigma,\omega)$ and four
$(\pi,\sigma,\omega,\rho)$ mesons exchange. In addition, it was shown that a
self-consistent treatment of the semi-relativistic nucleon wave function in
nuclear state has a great importance in calculations. The difference in
masses of $\sigma$ and $\omega$ mesons would not seriously change the main
aspect of the concept of relativistic or semi-relativistic interaction,
providing an average potential of cancelation of the repulsive meson
$(\omega)$ and the attractive meson $(\sigma)$ in conjunction with a weak
long range effect $(\pi)$. The work with $OBEP$ in Dirac-Hartree-Fock
equation gives a close relationship to other recent approaches, based upon
different formalisms which tended to support this direction. The nuclear
properties are being clear in our trail to include more two mesons to
describe the NN interaction through our potential. $SPED$ function has an
good ability to give us the better shapes of our potential and also better
values for energies. We hope that our potential represents a base for NN
interaction with different ranges of energies in following search. The ground
state energies for $^{2}H$ and $^{4}He$ nuclei are successfully determined
through this work, and gives us a hope to continue with more massive nuclei.
$QCD$ model is based on one-gluon exchange process besides the interaction of
instanton that supplemented the Confinement. The Cornell dressed potential
represents the interaction between quarks through the exchange of
pseudoscalar mesons (instantons) under controlling of one-gluon exchange
process. Our semi-relativistic hybrid model is encouraged for light nuclei,
and the instanton induced interaction caused to construct a link of
quark-quark interaction to the nucleon-nucleon interaction for small scale
around the hadron boundaries. The effect of adding the $QQ$ interaction on the ground state energies is ranged from $6.7$ for $^{2}H$ nucleus to $1.2$ for $^{4}He$ nucleus, this is small effect and it is expected to be vanished for massive nuclei.
\section{Conflict of Interest }
The authors declare that there is no conflict of interests
regarding the publication of this paper.
\appendixa
\par\noindent
We deals with the kinetic energy as a relative kinetic energy $T_{ij}$ which is related to the relative momentum $p_{ij}= \frac{1}{2}(p_{1}-p_{2})$ with the momentum of the first nucleon $p_{1}$ and momentum of the second nucleon $p_{2}$, and the center-of-mass momentum $p_{R}= p_{1}+p_{}$. Therefore, the relative kinetic energy has the formula,
\begin{eqnarray}
\nonumber                     T_{ij}&=& T_{i}-T_{c.m} \\
\nonumber                 &=& \sum_{i}\frac{p_{i}^{2}}{2m}- \frac{(\sum_{i}p_{i}){2}}{2mA} \\
\nonumber        &=& \sum_{i}\frac{p_{i}^{2}}{2m}-\frac{1}{2mA}[\sum_{i}p_{i}^{2}+\sum_{i<j}2p_{i}p_{j}]  \\
   \nonumber                   &=& \sum_{i}\frac{p_{i}^{2}}{2m}-\frac{1}{2mA}[\sum_{i}p_{i}^{2}+\sum_{i<j}(p_{i}^{2}+p_{j}^{2}-4p_{ij}^{2})]\\
     \nonumber                  &=& \sum_{i}\frac{p_{i}^{2}}{2m}-\frac{1}{2mA}[\sum_{i}p_{i}^{2}+(A-1)\sum_{i}p_{i}^{2}-4\sum_{i<j}p_{ij}^{2}]\\
     \nonumber                 &=&  \sum_{i}\frac{p_{i}^{2}}{2m}-\frac{1}{2mA}[A\sum_{i}p_{i}^{2}-4\sum_{i<j}p_{ij}^{2}]\\
                       &=&  \frac{2}{mA}\sum_{i<j}p_{ij}^{2}
                    \end{eqnarray}
Where $T_{i}$ is the kinetic energy of particles in the system, $T_{c.m}$ is the kinetic energy of the center-of-mass effect, $m$ is the mass of the nucleus and $A$ is the mass number of nucleus.

\appendixb
\par\noindent
The wave functions for two nucleons $i$ and $j$ have a form with Clebsch-Gordon coefficients.
\begin{eqnarray}
\nonumber&&\langle \phi_{\alpha}(r_{i})\phi_{\gamma}(r_{j})|=\\
\nonumber&&\sum_{m_{l_{\alpha}}m_{s_{\alpha}}}\sum_{m_{l_{\gamma}}m_{s_{\gamma}}}
   (l_{\alpha} s_{\alpha} m_{l_{\alpha}} m_{s_{\alpha}}|j_{\alpha} M_{\alpha})(l_{\gamma} s_{\gamma} m_{l_{\gamma}} m_{s_{\gamma}}|j_{\gamma} M_{\gamma})\\
 &&  \langle\phi_{n_{\alpha}l_{\alpha}m_{l_{\alpha}}}(r_{i})\phi_{n_{\gamma}l_{\gamma}m_{l_{\gamma}}}(r_{j})|
   \langle\chi_{m_{s_{\alpha}}}^{1/2}
   \chi_{m_{s_{\gamma}}}^{1/2}|\langle\hat{P}_{T_{\alpha}}\hat{P}_{T_{\gamma}}|
\end{eqnarray}
Where$(l)$ is the orbital angular momentum, $s_{\gamma}$ is the spin, the total angular momentum $j_{\alpha}=l_{\alpha}+ s_{\alpha}$ , $j_{\gamma}=l_{\gamma}+ s_{\gamma}$,
$ M_{\alpha}= m_{l_{\alpha}}+ m_{s_{\alpha}}$ in which $m_{l_{\alpha}}$ is the projection of orbital quantum number ,$m_{s_{\alpha}}$ is the projection of spin quantum number , $M_{\gamma}= m_{l_{\gamma}}+ m_{s_{\gamma}}$ and $\hat{P}_{T_{\alpha}}$ is the function of isotopic spin. The two wave functions are not connected and depend on $r_{i}$,$r_{j}$ so, the two wave functions need to be connected
\begin{eqnarray}
\nonumber&&\langle \phi_{\alpha}(r_{i})\phi_{\gamma}(r_{j})|=\\
\nonumber&&\sum_{m_{l_{\alpha}}m_{s_{\alpha}}}\sum_{m_{l_{\gamma}}m_{s_{\gamma}}}\sum_{\lambda\mu}
                                                      (l_{\alpha} s_{\alpha} m_{l_{\alpha}} m_{s_{\alpha}}|j_{\alpha} M_{\alpha})(l_{\gamma} s_{\gamma} m_{l_{\gamma}} m_{s_{\gamma}}|j_{\gamma} M_{\gamma})\\
                                                      &&(l_{\alpha}l_{\gamma}m_{l_{\alpha}}m_{l_{\gamma}}|\lambda\mu)
    \langle\phi_{n_{\alpha}l_{\alpha}m_{l_{\alpha}}}(r_{i})\phi_{n_{\gamma}l_{\gamma}m_{l_{\gamma}}}(r_{j})|
                                                     \langle\chi_{m_{s_{\alpha}}}^{1/2}\chi_{m_{s_{\gamma}}}^{1/2}|\langle\hat{P}_{T_{\alpha}}
                                                     \hat{P}_{T_{\gamma}}|
\end{eqnarray}
With $\lambda=l_{\alpha}+l_{\gamma}$ and $\mu=m_{l_{\alpha}}+m_{l_{\gamma}}$, we can change the special coordinates for each wave functions to become one wave, that depends on relative mass and center of mass.
\begin{eqnarray}
\nonumber&&\langle \phi_{\alpha}(r_{i})\phi_{\gamma}(r_{j})|=\\
\nonumber&&\sum_{m_{l_{\alpha}}m_{s_{\alpha}}}\sum_{m_{l_{\gamma}}m_{s_{\gamma}}}
    \sum_{\lambda\mu}\sum_{n l N L}(l_{\alpha} s_{\alpha} m_{l_{\alpha}} m_{s_{\alpha}}|
                                                      j_{\alpha} M_{\alpha})(l_{\gamma} s_{\gamma} m_{l_{\gamma}} m_{s_{\gamma}}|j_{\gamma} M_{\gamma})\\
\nonumber&&(l_{\alpha}l_{\gamma}m_{l_{\alpha}}m_{l_{\gamma}}|\lambda\mu)
    \langle n_{\alpha}l_{\alpha}n_{\gamma}l_{\gamma}|NL nl\rangle(s_{\alpha}s_{\gamma}m_{s_{\alpha}}m_{s_{\gamma}}|S M_{s})
    \langle\phi_{{NL} {nl}}(r,R)|\\
    &&\langle\chi_{m_{s_{\alpha}}}^{1/2}\chi_{m_{s_{\gamma}}}^{1/2}|
\langle\hat{P}_{T_{\alpha}}\hat{P}_{T_{\gamma}}|
\end{eqnarray}
Where $\langle n_{\alpha}l_{\alpha}n_{\gamma}l_{\gamma}|NL nl\rangle$ is the Talmi-Moshinsky bracket , $NL$ is total center of mass , $nl$ is  total relative. The wave function $\phi_{{NL} {nl}}(r,R)$ can be spitted in to the form
\begin{eqnarray}
\nonumber&&\langle \phi_{\alpha}(r_{i})\phi_{\gamma}(r_{j})|=\\
\nonumber&&\sum_{m_{l_{\alpha}}m_{s_{\alpha}}}\sum_{m_{l_{\gamma}}m_{s_{\gamma}}}\sum_{JM}
    \sum_{\lambda\mu}\sum_{n l N L}\sum_{m M}(l_{\alpha} s_{\alpha} m_{l_{\alpha}} m_{s_{\alpha}}|
                                                      j_{\alpha} M_{\alpha})(l_{\gamma} s_{\gamma} m_{l_{\gamma}} m_{s_{\gamma}}|j_{\gamma} M_{\gamma})\\
 \nonumber                                                     &&(l_{\alpha}l_{\gamma}m_{l_{\alpha}}m_{l_{\gamma}}|\lambda\mu)
    \langle n_{\alpha}l_{\alpha}n_{\gamma}l_{\gamma}|NL nl\rangle(s_{\alpha}s_{\gamma}m_{s_{\alpha}}m_{s_{\gamma}}|S M_{s})(l S m_{l} m_{S}|J M)\\
    &&(L l M m|\lambda\mu)
    \langle\phi_{NLM}(R)\phi_{nlm}(r)|
    \langle\chi_{m_{s_{\alpha}}}^{1/2}\chi_{m_{s_{\gamma}}}^{1/2}|\langle\hat{P}_{T_{\alpha}}\hat{P}_{T_{\gamma}}|
\end{eqnarray}
As $L$ gives the total orbital quantum number in center of mass , $l$ gives the total orbital quantum number in relative coordinates and $S=s_{i}+s_{j}$ is the total spin.
Relative to the spin functions and isospin functions to be connected ,we have to use them as following.

\begin{eqnarray}
\nonumber&&\langle \phi_{\alpha}(r_{i})\phi_{\gamma}(r_{j})|=\\
\nonumber&&\sum_{m_{l_{\alpha}}m_{s_{\alpha}}}\sum_{m_{l_{\gamma}}m_{s_{\gamma}}}\sum_{JM}
    \sum_{\lambda\mu}\sum_{n l N L}\sum_{m M}\sum_{s m_{s}}\sum_{T}(l_{\alpha} s_{\alpha} m_{l_{\alpha}} m_{s_{\alpha}}|
                                                      j_{\alpha} M_{\alpha})(l_{\gamma} s_{\gamma} m_{l_{\gamma}} m_{s_{\gamma}}|j_{\gamma} M_{\gamma})\\
  \nonumber        &&(l_{\alpha}l_{\gamma}m_{l_{\alpha}}m_{l_{\gamma}}|\lambda\mu)
\langle n_{\alpha}l_{\alpha}n_{\gamma}l_{\gamma}|NL nl\rangle(l S m_{l} m_{S}|J M)(L l M m|\lambda\mu)\\
\nonumber && (s_{\alpha}s_{\gamma}m_{s_{\alpha}}m_{s_{\gamma}}|S M_{s})
 (\chi_{\alpha}\chi_{\gamma}T_{\alpha}T_{\gamma}|T M_{T})\langle\phi_{NLM}(R)\phi_{nlm}(r)\\
 &&\langle\chi_{m_{s_{\alpha}}}^{1/2}(i,j)|\langle\hat{P}_{T}(i,j)|
\end{eqnarray}
 With the isotopic spin $T_{proton}=\frac{-1}{2}$ and $T_{neutron}=\frac{1}{2}$.

\end{document}